\documentclass[useAMS,usenatbib]{mn2e}

\usepackage {graphicx}

\title[Influence of outflows on luminosity fluctuations]
{The influence of outflows on the $1/f$-like luminosity fluctuations}

\author[D.-B. Lin, W.-M. Gu, T. Liu, and J.-F. Lu]
{Da-Bin Lin, Wei-Min Gu\thanks{E-mail:
guwm@xmu.edu.cn}, Tong Liu, and Ju-Fu Lu\\
Department of Physics and Institute of Theoretical Physics and
Astrophysics, Xiamen University, Xiamen, Fujian 361005, China}

\begin{document}

\date{}

\maketitle

\begin{abstract}
In accretion systems, outflows may have significant influence on
the luminosity fluctuations.
In this paper, following the Lyubarskii's general scheme, we revisit
the power spectral density of luminosity fluctuations
by taking into account the role of outflows.
Our analysis is based on the assumption that the coupling
between the local outflow and inflow
is weak on the accretion rate fluctuations.
We find that, for the inflow mass accretion rate $\dot M \propto r^{s}$,
the power spectrum of flicker noise component will present
a power-law distribution
$p(f) \propto f^{-(1+4s/3)}$ for advection-dominated flows.
We also obtain descriptions of $p(f)$ for
both standard thin discs and neutrino-cooled discs,
which show that the power-law index of a neutrino-cooled
disc is generally larger than that of a photon-cooled disc.
Furthermore, the obtained relationship between $p(f)$ and $s$
indicates the possibility of evaluating the strength of outflows
by the power spectrum in X-ray binaries and gamma-ray bursts.
In addition, we discuss the possible influence of 
the outflow-inflow coupling on our results.
\end{abstract}


\section{Introduction}
The emission of Galactic Black Hole Binaries (BHBs) and active
galactic nuclei (AGN) displays a significant aperiodic variability
on a broad range of time-scales.
The Power Spectral Density (PSD) of such variability is generally
modeled with a power law, $p(f) \propto f^{-\beta}$,
where $p(f)$ is the power at frequency $f$, and the power-law index
${\beta}$ keeps a constant in a certain range of $f$,
but changes among different
ranges. At high frequencies,
the PSDs of both BHBs and AGN present a steep slope 
with $\beta \sim 2$. On the contrary, below a break frequency,
typically at a few Hz for BHBs, they flatten to a slope with $\beta \sim 1$,
representing the flicker noise (see \citealp{King04} and references therein).

Several models have been proposed in order to understand this nearly
featureless character of power spectra.
The so-called ``shot noise models" (\citealp{Terrell72}) attempted to describe
the light curves as a series of independent overlapping shots with 
specific time-scales, amplitudes, and occurence rates.
Due to lack of physical picture in this scenario, 
various physically motivated ideas have been put forward subsequently,
such as the fluctuations of hydrodynamic or magnetohydrodynamic turbulence
(\citealp{Nowak95}; \citealp{Hawley01}),
magnetic flares or density fluctuation in the corona
(\citealp{Galeev79}; \citealp{Poutanen99}; \citealp{Goosmann06}; 
\citealp{Kawanaka08}), and Lyubarskii's general scheme
(\citealp{Lyubarskii97}; \citealp{King04}).
In the Lyubarskii's scheme,
it was noted that any variation of 
accretion rate, which is caused by small amplitude variations
in the viscosity, would induce a variation in the accretion 
rate at the inner radius of the disc, where most
of the energy is released.
Moreover, observations showed
that the variability is non-linear and the rms variability 
is proportional to the average flux over a wide 
range of time-scales (e.g., \citealp{Uttley01}; 
\citealp{Uttley05}; \citealp{Gleissner04}).
It indicates
that the short time-scale variations
are modulated by the longer time-scales, which
favors the Lyubarskii's scheme.

However, the observed power spectra often deviate from the form $f^{-1}$.
For example, the PSD of Cyg X-1 is well described with the form $f^{-1}$
in the soft state, however, exhibits the form $f^{-1.3}$
in the hard state (\citealp{Gilfanov10}). 
In particular, it is shown that the power-law index
is around $0.8-1.3$ both in the soft state of
BHBs and in narrow-line Seyfert 1 galaxies
(\citealp{Janiuk07}).
Such a dispersion of the power-law index reveals that
there must exist some other mechanism.
A radius-dependent amplitudes  
of $\alpha$ fluctuations may help to 
alleviate the discrepancy between theories and observations.
However, it remains unclear why the fluctuations of $\alpha$ should
have a strong radius-dependent form.

In the present paper, we will take into account another mechanism, outflows,
which is a popular phenomenon in accretion systems and has strong
observational evidence.
One of the best examples comes from Sgr A*, whose center 
harbors a supermassive black hole surrounded by an accretion flow that is 
likely to be in the form of the advection-dominated accretion flow (ADAF, \citealp{Narayan94}). 
Radio polarization observations constrain
the accretion rate in the innermost region is nearly two orders of
magnitude lower than that measured at the Bondi radius
(e.g., \citealp{Marrone06}), 
which indicates that intense outflows may present in this system.
Besides, the absorption lines from highly ionized elements, which have been 
detected in the X-ray spectrum of some microquasars such as GRO J1655-40 (\citealp{Ueda98}; 
\citealp{Yamaoka01}; \citealp{Miller06}), GRS 1915+105 (\citealp{Kotani00}; 
\citealp{Lee02}) and Atoll sources
(see e.g. the review by \citealp{Diaz Trigo06} and references therein),
also indicate the existence of outflows. 
On the other hand, \citet{Jiao11} found that outflows generally
exist in accretion discs no matter that the flow is
advection-dominated such as the slim disc
(\citealp{Abramowicz88}) and the ADAF, or is radiation-dominated such as
the standard thin disc (\citealp{Shakura73}).
In particular, for the three types of advection-dominated flows: ADAFs
(gas internal energy dominant), slim discs
(trapped photon energy dominant), and
hyper-accretion discs (trapped neutrino energy dominant),
outflows may be significantly strong due to positive Bernoulli parameters (e.g., \citealp{Narayan97}; \citealp{Liu11}) or
the large radiation pressure (e.g., \citealp{Gu07}).
Furthermore, outflows have generally been found in many
simulation works (e.g., \citealp{Ohsuga11} and references therein).

In the Lyubarskii's scheme, the power spectrum of luminosity fluctuations
is sensitive to the varying mass accretion rate, thus we expect that
outflows may have significant effects on the power-law index.
The paper is organized as follows. 
In Section 2, we investigate the fluctuation power spectrum under
a radius-dependent accretion rate following
the method of Lyubarskii.
The potential application of our results to observations
is discussed in Section 3.

\section{Luminosity fluctuations with a radius-dependent accretion rate}

\subsection{Evolution equations of the fluctuations}

In the present work, following the Lyubarskii's general scheme,
we revisit the power spectral density of luminosity fluctuations
by considering a radius-dependent accretion rate.
The relevant processes are the conservation of mass and 
angular momentum. With a radius-dependent accretion rate, 
these two processes are described as follows:
\begin{equation}\label{mass_conservation}
\frac{\partial \Sigma }{\partial t} = \frac{1}{2\pi{r}}(\frac{\partial 
\dot {M}}{\partial r} - \Phi ),
\end{equation}
\begin{equation}\label{Angular_conservation}
\frac{\partial (\dot {M}{\Omega}r^2)}{\partial r} = - \frac{\partial 
}{\partial r}(2\pi r^2{\rm T}_{r\varphi} ) + {\Phi} l_{\Phi},
\end{equation}
where $\Sigma $, $\dot {M}$, $\Omega$ and ${\rm T}_{r\varphi} $ are the 
surface density, accretion rate, angular velocity and $r\varphi$ 
component of the stress tensor at the radius $r$, respectively. 
${\Phi} $ is the change of accretion rate over
$r$ in the stationary accretion state, i.e. ${\Phi}={\partial
\dot {M}}/{\partial r}$, and $l_{\Phi} $ is the angular 
momentum of ${\Phi}$.

With ${\Omega}r^2 = l_{\rm in} $, $F(=-{\rm T}_{r\varphi} r^2)= 0$ at the inner 
radius $r_{\rm in} $, we can obtain
\begin{eqnarray*}
F&=&\frac{1}{2\pi}[\dot {M}{\Omega}r^2 - \dot {M}_{\rm in} l_{\rm in}-\int_{r_{\rm in} }^r {{\Phi} l_{\Phi} dr}] \\
&=&\frac{\dot {M}{\Omega}r^2}{2\pi}[1 - (\dot {M}_{\rm in} l_{\rm in}+\int_{r_{\rm in} }^r {{\Phi} l_{\Phi} dr})/(\dot {M}{\Omega}r^2)].
\end{eqnarray*}
With the following definitions:
\begin{equation}
h = \Omega r^2,
\end{equation}
\begin{equation}\label{Q}
Q = 1 - (\dot {M}_{\rm in} l_{\rm in}+\int_{r_{\rm in} }^r {{\Phi} l_{\Phi} dr})/(\dot {M}{\Omega}r^2),
\end{equation}
the above description of $F$ can be simplified as
\[
F = \frac{\dot {M}h}{2\pi }Q.
\]
For standard thin discs and ADAFs,
$\Omega \propto \Omega_{\rm K} = \sqrt {GM / r^3}$, and thus $h 
\propto r^{1 / 2}$. We assume $h=b\sqrt {GMr}$ in this paper,
where $b$ is a constant.
If the accretion rate has weak dependence on the radius
and $l_{\Phi} \propto {\Omega}r^2 \propto \sqrt{r}$, 
the value of $Q$ would remain nearly constant. 
In order to simplify the problem, we assume $Q$ to be constant in our analysis.
The validity of this assumption is discussed in Appendix~A.

With $\chi =(1/2\pi)\int {{\Phi} (l_{\Phi} - \Omega r^2)dr}$,
equation (\ref{Angular_conservation}) becomes
\begin{equation}\label{AccretionR01}
\dot {M} = - \frac{\partial [2\pi (F + \chi) ]
}{\partial h} .
\end{equation}
Substituting the above equation into the equation (1), we have
\begin{equation}\label{perturbation03}
\frac{\partial \Sigma }{\partial t} = \frac{b^4(GM)^2}{2h^3}\frac{\partial ^2(F 
+ \chi )}{\partial h^2} - \frac{{b^2GM}}{2{\pi}h^2}{\Phi}.
\end{equation}
The relationship between $\Sigma$ and $F$ can be deduced with 
$\alpha $-prescription and the local balance between the heating
and the cooling
in the disc. The formula is similar to that with a constant
accretion rate 
(\citealp{Filipov84}; \citealp{Lyubarskii87}; \citealp{Narayan94}), i.e.
\begin{equation}\label{relationship04}
\Sigma = \frac{b^4(GM)^2F^{1 - m}}{2(1 - m)Dh^{3 - n}},
\end{equation}
where the exponents $m$ and $n$ are determined by the disc model,
which are discussed in \S2.2, and $D$ is a function of $\alpha$.
It should be noted that the analysis presented here is not strict
for ADAFs. A stringent analysis for the ADAFs may refer to Appendix~B.

Substituting equation (\ref{relationship04}) into equation (\ref{perturbation03}), we have 
\begin{eqnarray}\label{InitialP02}
\frac{\partial F}{\partial t} = \frac{2DF^m}{b^4(GM)^2h^{n - 3}}
[\frac{b^4(GM)^2}{2h^3}\frac{\partial ^2(F + \chi )}{\partial h^2} - \nonumber \\
 \frac{{b^2GM}}{2{\pi}h^2}{\Phi}] + \frac{F}{(1 - m)D}\frac{\partial D}{\partial t}.
\end{eqnarray}
By assuming $\alpha = \alpha _0 [1 + \overline{\beta} (t,r)]$, 
where $\overline{\beta}(t,r) (\ll 1)$ is de-correlated at
different radial scales and its correlation time-scale
is of the order of the local viscous time-scale, 
the disturbed quantities $D$ and $F$ are
\[
D = D_0 (1 + \eta \overline{\beta} ),\;\;\;\;F = (\frac{\dot {M}_0 h}{2\pi })Q + \psi, 
\]
where $\eta$ is defined as $D \propto {\alpha}^{\eta}$ and the subscript 
$0$ denotes unperturbed quantities.
Substituting the above equations into equation (\ref{InitialP02}), and
including the following stationary condition
\[
\frac{b^4(GM)^2}{2h^3}\frac{\partial ^2}{\partial h^2}(\frac{\dot {M}_0 h}{2\pi 
}Q + \chi ) - \frac{{b^2GM}}{2{\pi}h^2}{\Phi} = 0,
\]
we have,
\begin{equation}
\frac{\partial \psi }{\partial t} = \frac{D_0 Q^m}{h^{n - m}}(\frac{\dot 
{M}_0 }{2\pi })^m\frac{\partial ^2\psi }{\partial h^2} + \frac{\dot {M}_0 
hQ}{2\pi (1 - m)}\frac{\partial \eta \overline{\beta} }{\partial t}.
\end{equation}
Here, we assume that the coupling between the local outflow and inflow
is weak on the accretion rate fluctuations.
A simple discussion on this issue is presented in \S2.4.

\subsection{The analysis of luminosity fluctuations}

Assuming the radius-dependent accretion rate is
\begin{equation}\label{AccretionRate}
\dot {M}_0 \; = \;\dot {M}_{0,\;\rm in} (\frac{r}{r_{\rm in} })^{s},
\end{equation}
where $\dot {M}_{0,\;\rm in}$ is the accretion rate at the inner radius $r_{\rm in}$,
we have
\[
\frac{\partial \psi }{\partial t} = \frac{C}{h^{n - m - 2{s} m}}(\frac{\dot 
{M}_{0,\;{\rm in}} }{2\pi })^m\frac{\partial ^2\psi }{\partial h^2} 
\]
\begin{equation}\label{perturbation06}
\quad\quad\quad\quad
+ \frac{\dot {M}_{0,\;{\rm in}} hQ\eta }{2\pi (1 - m)}\frac{\partial }{\partial t}
[\overline{\beta }(\frac{h^{2}}{b^2GMr_{{\rm in}}})^{s}]
\end{equation}
and $C = D_0 Q^m/(GMr_{\rm in}b^2)^{{s} m}$. 
This is a linear diffusion equation with a radius-dependent accretion rate.
For ${s} = 0$, i.e. a constant accretion rate in the system, we have $Q=1$
and equation (\ref{perturbation06}) is reduced to
\begin{equation}\label{Lyubarskii}
\frac{\partial \psi }{\partial t} = \frac{D_0 }{h^{n - m}}(\frac{\dot 
{M}_{0,\;{\rm in}} }{2\pi })^m\frac{\partial ^2\psi }{\partial h^2} + \frac{\dot 
{M}_{0,\;{\rm in}} h\eta }{2\pi (1 - m)}\frac{\partial \overline{\beta} }{\partial t}, 
\end{equation}
which is the exact form of equation~(9) in \citet{Lyubarskii97}.

In general, the solution of equation (\ref{perturbation06}) is (\citealp{Lyubarskii97}, \citealp{Lynden-Bell74})
\begin{eqnarray*}
\psi (t,x) - \psi (0,x) =
\frac{\dot {M}_{0,\;{\rm in}} \eta Q\kappa ^2x^l}{4\pi(1 - m)}
\int_0^t {dt'} \int_0^\infty {dx_1 } \frac{x_1^{l + 1} }{t - t'}\quad\quad\\
\quad\exp [-\frac{(x^2 + x_1^2 )\kappa ^2}{4(t - t')}]I_l [\frac{\kappa ^2xx_1 }{2(t - 
t')}]\frac{\partial }{\partial t'}[\overline{\beta} (t',x_1 ) {\frac{x_1^{4ls}}{(b^2GMr_{\rm in})^{s}}}],
\end{eqnarray*}
where
\begin{equation}
l = \frac{1}{2 + n - m - 2{s} m}\;,\,
\end{equation}
\[
x = h^{1/{2l}},\,
4(\frac{l}{\kappa })^2 = C(\frac{\dot {M}_{0,\;{\rm in}} }{2\pi 
})^m.
\]
From equation (2), the accretion rate $\dot{M}(t,x)$ is
\[
\dot{M}(t,x) - \dot{M}(0,x)=
\]
\[
\;\;\;\;\int_0^t {\int_0^\infty {G(t,x;t',x_1 )\frac{\partial 
}{\partial t'}[\overline{\beta} (t',x_1 ) {\frac{x_1^{4ls}}{(b^2GMr_{\rm in})^{s}}}]} } dt'dx_1,
\]
where
\begin{eqnarray*}
G(t,x;t',x_1 ) = \frac{\eta Q\kappa ^4x^{1 - l}x_1^{l + 1} }{8l(1 - m)(t - 
t')^2} \exp [ - \frac{(x^2 + x_1^2 )\kappa ^2}{4(t - t')}] \\
\{x_1 I_{l - 1} 
[\frac{\kappa ^2xx_1 }{2(t - t')}] - xI_l [\frac{\kappa ^2xx_1 }{2(t - 
t')}]\} ,
\end{eqnarray*}
and $I_\nu (z)$ is the Bessel function of the imaginary argument.
Based on the above equation, the power spectrum of $\dot{M}(t,x)$
can be obtained following a complex calculation as presented in Section 4
of \citet{Lyubarskii97}.
Here, we directly present the result of the power spectrum
and focus on the effects of outflows on the luminosity fluctuations.
If $\sqrt { < (\overline{\beta} )^2 > } \propto \;r^\xi$, the power 
spectrum $p(f)$ of $\dot{M}(t,x)$ is
\begin{equation}
p(f) \propto f^{ - [1 + 4l(\xi + {s} )]},
\end{equation}
which indicates the power-law index $\beta = 1 + 4l(\xi + {s} )$.
In this work we ignore $\alpha$ fluctuations, i.e.
$\xi=0$, and then the expression of $\beta$ will be reduced to
\begin{equation}
\beta = 1 + 4 l s .
\end{equation}

For an advection-dominated flow, we have $m = 0$, $n = 1$, and $D\propto \alpha$
(\citealp{Narayan94}) for equation (7), thus equation (13) indicates
$l = 1/3$ and therefore $\beta$ can be simplified as
\begin{equation}
\label{ADAF}
\beta = 1 + \frac{4}{3} s.
\end{equation}
We would like to stress that the above formula should be valid
for all the three types of advection-dominated flows mentioned in
the first section.

For a standard thin disc, it is well-known that there exist three
regions according to different dominant mechanisms for opacity
and pressure.
We have
$m = 0.3$, $n = 0.8$, and $D \propto \alpha^{0.8}$ for the outer region
($p \sim p_{\rm gas}$, $\kappa \sim \kappa_{\rm ff}$),
$m = 0.4$, $n = 1.2$, and $D \propto \alpha^{0.8}$ for the middle region
($p \sim p_{\rm gas}$, $\kappa \sim \kappa_{\rm es}$),
and $m = 2$, $n = 7$, and $D \propto \alpha$ for the inner region
($p \sim p_{\rm rad}$, $\kappa \sim \kappa_{\rm es}$),
thus we easily obtain the following from equations (13) and (15):
\begin{equation}
\label{Outer}
\beta = 1 + \frac{40}{25-6s}{s}
\end{equation}
for the outer region,
\begin{equation}
\label{Middle}
\beta = 1 + \frac{10}{7-2s}{s}
\end{equation}
for the middle region, and
\begin{equation}  \label{Inner}
\beta = 1 + \frac{4}{7-4s}{s}
\end{equation}
for the inner region.
We would point out that outflows may be negligible
in standard thin discs expect for the inner region, which is
radiation pressure dominated and may suffer the thermal
instability. Outflows in the inner region may be significantly stronger
than that in the other two regions due to
the thermal instability. However, it remains under controversy
whether the inner region is indeed thermally unstable or not
(e.g., \citealp{Hirose09}), and many
mechanisms have recently been proposed to suppress the instability
(e.g., \citealp{Zheng11}; \citealp{Lin11}; \citealp{Ciesielski11}).

Apart from the above mentioned photon-radiation-dominated flows,
neutrino-dominated accretion flows (NDAFs) have also been
widely studied (e.g., \citealp{Popham99}; \citealp{Narayan01}; \citealp{Gu06};
\citealp{Chen07}; \citealp{Liu07}),
which may account for the central
engine of gamma-ray bursts (GRBs, e.g., \citealp{Narayan92}).
The outer region of NDAFs may be advection-dominated since
neutrino cooling cannot balance the viscous heating due to
low temperature and density. This region can be regarded as
an extension of slim discs (e.g., \citealp{Liu08}).
On the other hand, for high mass
accretion rates such as $\dot m \ga 1 M_{\sun}$ s$^{-1}$,
the inner region may also become
advection dominant due to the large optical depth for neutrinos
(e.g., \citealp{DiMatteo02}; \citealp{Gu06}). Outflows may occur
in NDAFs in particular for these two advection-dominated regions owing
to positive Bernoulli parameters (e.g., \citealp{Liu11}).

For NDAFs, with the balance between the
cooling rate by pair capture and the energy dissipation rate
per unit volume (\citealp{Popham99}, Eqs.~5.4-5.5),
we have $m = 0$, $n = -0.4$, and $D \propto \alpha^{1.2}$, 
thus equations (13) and (15) directly give the form of $\beta$:
\begin{equation}
\label{GRB}
\beta = 1 + \frac{5}{2}{s}.
\end{equation}

\begin{figure}
\begin{center}
{\includegraphics[width=3.00in,height=2.70in]{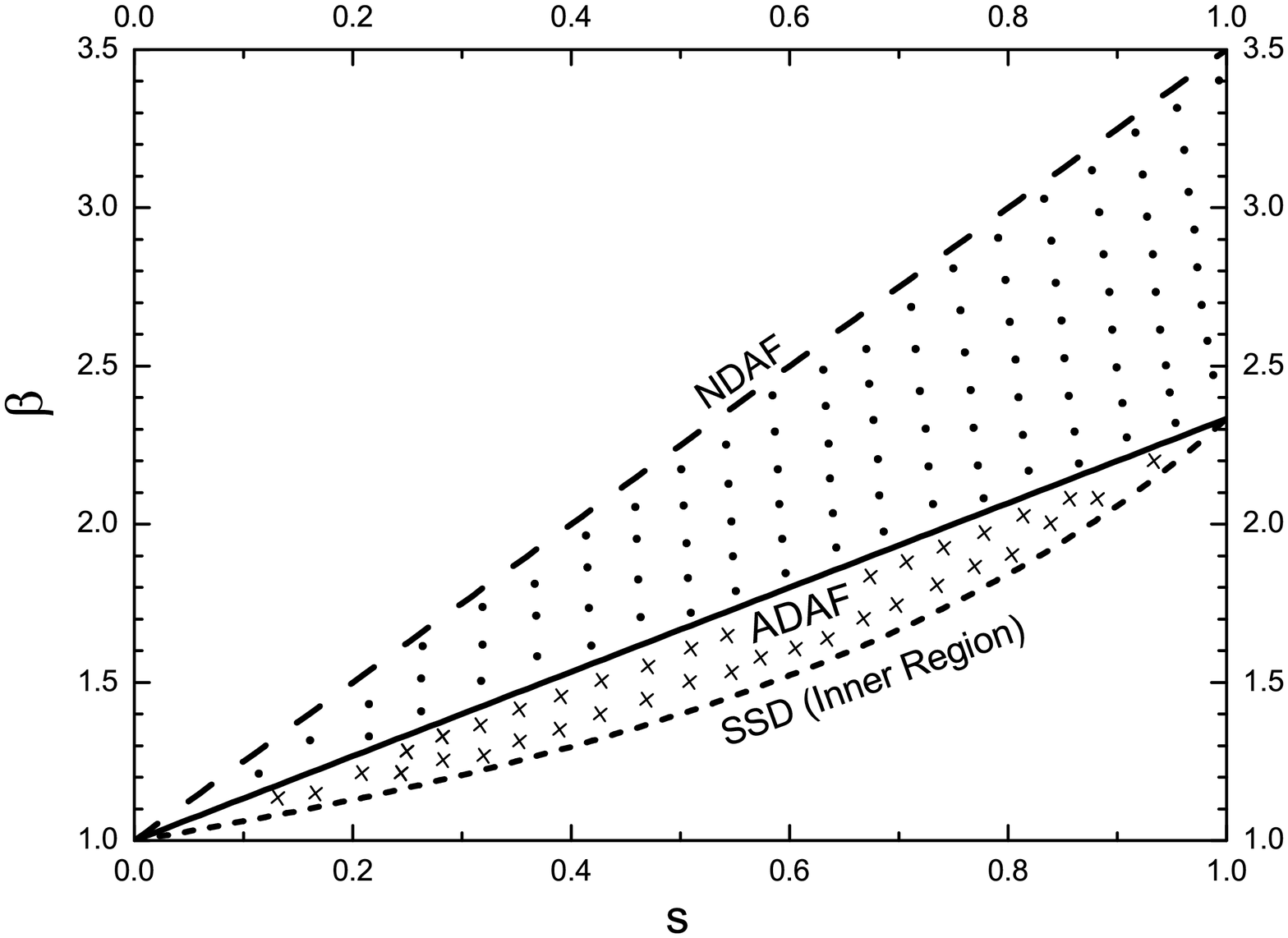}}
\caption{Variation of $\beta$ with $s$ for three types of accretion models:
radiation-pressure-supported
standard thin discs (the short dashed line), ADAFs (the solid line), and
NDAFs (the long dashed line). The region of photon-cooled discs and that
of neutrino-cooled discs are marked by
the cross symbol and the dot symbol, respectively.
\label{fig1}}
\end{center}
\end{figure}

The relationship between $\beta$ and $s$ for all the mentioned accretion
models are presented in Figure~\ref{fig1}, where
the short dashed, solid, and long dashed lines correspond to
the inner region of standard thin discs, ADAFs, and NDAFs, respectively.
The region with cross symbols represents photon-cooled discs
(cooling dominated either by photon radiation
or by advection of photon or gas energy),
whereas the region with dot symbols represents neutrino-cooled discs
(cooling dominated either by neutrino radiation or by advection
of neutrino energy).
The figure clearly indicates that, for comparable $s$,
a neutrino-cooled disc generally has a larger $\beta$ than a
photon-cooled disc. In other words, $\beta$ in
a GRB system is generally larger than that in a BHB system.

\subsection{A simple analysis}
The dependence of $\beta$ on $s$ can be well understood
with equations (\ref{AccretionR01}) and (\ref{Lyubarskii}).
Following these two equations, the amplitude of $\delta \dot{M}(r)$, 
which is the accretion rate variation produced at a radius $r$, 
is proportional to the accretion rate at that radius, 
i.e., $\delta \dot{M}(r) \propto \dot{M}(r)$.
Since variations of the accretion rate in the inner region 
may be represented as a sum of independent accretion rate 
variations produced at different radii (\citealp{Lyubarskii97}),
the power spectrum for the flow with $\dot {M} \propto r^{s}$ can
be written as 
\begin{equation}\label{21}
p(f) \propto \dot{M}^2 f^{-1} \propto r^{2s} f^{-1},
\end{equation}
where $\dot{M}$ is the accretion rate at the radius $r$ which
contributes to the fluctuation at the frequency $f$,
and the term $f^{-1}$ is taken from \citet{Lyubarskii97} for
the constant accretion rate case.

Taking ADAFs as an example, we have $f \propto 1/t_{\rm vis}$
and $t_{\rm vis}=(r/H)^2/(\alpha\Omega)\propto r^{3/2}$,
i.e., $r \propto f^{-2/3}$.
Equation (\ref{21}) is therefore simplified as
\[
p(f) \propto f^{-(1 + \frac{4}{3}{s})},
\]
which is exactly the same form as equation (16). 
For other disc models, such as standard thin discs and NDAFs,
the results can be understood in the same way.

\subsection{Possible influence of the outflow-inflow coupling}

As mentioned in \S2.1,
our results are based on the assumption that the coupling between
the local outflow and inflow is weak on the accretion rate fluctuations,
i.e., the response of outflows to the inflow fluctuations is weak.
In principle, the outflow should have
fluctuations related to the variations of the local inflow accretion rate.
However, the mechanism for outflows may be complicated, 
and therefore the fluctuations of the outflow may not be determined simply
by that of the local inflow.
For example, radiation from the inner region of a disc can heat up
the materials in the outer region and thus outflows can be produced
(e.g., \citealp{Begelman83}; \citealp{Metzger08}).
In such case the outflow may be more relevant to the inner disc
rather than the outer region where it occurs. 

Here, we make a simple discussion on the possible influence
of the local outflow-inflow coupling as follows.
If the coupling exists,
the amplitude of accretion rate fluctuation may vary 
while it propagates into the inner region.
With the assumption that $\delta \dot{M}(r)|_{r_1}$ is the accretion rate
variation at the radius $r_1$ induced by $\delta \dot{M}(r)$, we introduce 
a factor $D(r, r_1)$ to generally describe the change of the above
amplitude owing to the coupling effects, i.e.,
$\delta \dot{M}(r)|_{r_1} = D(r, r_1) \delta \dot{M}(r)$.
It is easy to find the form of $D(r, r_1)$ in the following two situations:
[1], if the coupling is negligible, the amplitude of $\delta \dot{M}(r)$
will keep unchanged during its propagation, 
i.e., $\delta \dot{M}(r)|_{r_1}=\delta \dot{M}(r)$, thus $D(r, r_1)=1$; 
[2], if the local outflow is in strong coupling with the local inflow,
the relative amplitude $\delta \dot{M}(r)|_{r_1}/\dot{M}(r_1)$
will keep unchanged for varying ${r_1}$,
i.e., $D(r, r_1) = \dot{M}(r_1)/\dot{M}(r)$.
Obviously, our results are under the former situation.
On the contrary, for the latter one,
the power spectrum of accretion rate variations
at the inner radius $r_{\rm in}$ can be expressed as 
\[
p(f) \propto D(r, r_{\rm in})^2 \dot{M}^2 f^{-1}
= \dot{M}_{\rm in}^2 f^{-1} \propto f^{-1},
\]
which is the same form as the result with a constant accretion rate.
(e.g., \citealp{Lyubarskii97}).
The detailed prescription of $D(r, r_1)$ is, however, beyond the scope 
of the present work due to the complexity of outflows.
Nevertheless, we can expect that a real flow may exist
between situations [1] and [2], and therefore
the value of $\beta$ may be located between unity and
the results presented in \S2.2.

\section{Summary and Discussion}
In this paper, we evaluate the effects of outflows 
on the luminosity fluctuations with the Lyubarskii's general scheme.
With a radius-dependent accretion rate $\dot M \propto r^{s}$, 
the power spectrum of the luminosity fluctuations is 
$p(f) \propto f^{-\beta}$, where the value of $\beta$
varies with the disc structure.
By assuming that the coupling between the local outflow
and inflow is weak on the accretion rate fluctuations,
we obtain the following
explicit expressions of $\beta$ for different disc models: 
$\beta = 1 + 4s/3$ for advection-dominated discs,
$\beta = 1 + 40s/(25-6s)$ for the outer region of standard thin discs, 
$\beta = 1 + {10s}/(7-2s)$ for the middle region, 
$\beta = 1 + {4}{s}/({7-4s})$ for the inner region,
and $\beta = 1 + 5s/2$ for NDAFs.
The above expressions imply that $\beta$ in a GRB
is generally larger than that in a BHB for comparable $s$. 
The expressions of $\beta$
indicate the possibility of evaluating the strength of outflows
by the power spectrum in X-ray binaries and GRBs.
In addition, if the coupling is not negligible, the value of $\beta$
will probably be located between unity and the value presented
in the above expressions.

In both BHBs and AGN, ADAFs are usually adopted to describe
the quiescent state, the low/hard state, 
and the corona which lies above a cold disc.
ADAFs may produce significant outflows, 
and therefore the power spectrum can deviate from $f^{-1}$ based on
the present analysis.
The exact value of ${s}$ is, however, difficult to estimate from
the theoretical point of view, except for
the general constraint $0 < s < 1$ (\citealp{Narayan08}).
On the other hand, some observations
indicate $s \sim 0.3$ (\citealp{Yuan03}; \citealp{Zhang10}).
Taking this value, we obtain $\beta=1.4$ for ADAFs,
which is close to $1.3$,
the power-law index of PSDs presented in the low mass X-ray binary 
systems (\citealp{Gilfanov05}). The quantitative difference
may be relevant to the coupling between the outflow and inflow
as discussed in \S2.4.

If there is only outflow that operates in the accreting system, 
$s$ should be positive.
However, $s$ can also be negative due to the evaporation mechanism of
a cold disc.
For the accreting black hole in BHBs, the observed
power-law components in the X-ray spectra are generally
attributed to hot, tenuous plasmas, namely accretion disc
coronae. Due to the high temperature in the corona, the interaction
between the disc and corona would lead to mass
evaporating from the disc to the corona (\citealp{Meyer00}; \citealp{Spruit02}). 
In this case, the value of $s$ for the corona should be 
negative if outflows are not strong, and therefore
it is quite possible for $\beta$ to be less than unity.
Consequently, in this scenario
$s$ for the underneath cold disc should be positive.

\section*{Acknowledgments}
    
We thank Feng Yuan, Wen-Fei Yu, and Shan-Shan Weng for beneficial discussion, 
and the referee for helpful comments.
This work was supported by the National Basic Research Program of China
under grant 2009CB824800, and the National Natural Science Foundation
of China under grants 10833002, 11073015, and 11103015.

\appendix

\section[]{The validity of the assumption of constant $Q$}

In this section, we analyze the validity of the assumption of constant $Q$.

With a radius-dependent accretion rate described in equation (\ref{AccretionRate}), we have 
\begin{displaymath}
{\Phi}={{s}} \frac{\dot{M}_{0,\rm in}}{r_{\rm in}}(\frac{r}{r_{\rm in}})^{({s}-1)},
\end{displaymath}
then equation (\ref{Q}) can be expressed as
\[
Q = 1 - {{{{\dot M}_{\rm in}}{l_{\rm in}} + {s} \int_{{r_{\rm in}}}^r \frac{\dot{M}_{0,\rm in}}
{r_{\rm in}}(\frac{r}{r_{\rm in}})^{({s}-1)}{l_{\Phi}}dr} \over {\dot M\Omega {r^2}}}.
\]
We assume that the specific angular momentum corresponding 
to $\Phi$ is proportional to that of the gas in the disc, i.e.
\[
l_{\Phi} \propto \Omega {r^2} \propto {r^{1/2}}.
\]
This is the case for thermal energy driving outflows,
magnetic field centrifugal accelerating wind (\citealp{Mayer06}) and disc evaporation model.

For the general situation (${s}  \neq -1/2$), we have 
\[
Q = 1 - {{{{\dot M}_{\rm in}}{l_{\rm in}} + {{s}  \over {{s}  + 0.5}}
\left[ {\dot M{l_{{\mathop{\Phi}} }}} \right]_{{r_{\rm in}}}^r} \over {\dot M\Omega {r^2}}}
\]
\[ 
\;\;\;\;= 1 - \frac{s}{s+ 0.5}\frac{l_{\Phi}} 
{\Omega {r^2}} - ({l_{\rm in}} - {{s}  \over {{s}  + 0.5}}{l_{{\mathop{\Phi}} ,{\rm in}}})
{({{{r_{\rm in}}} \over r})^{s} }{1 \over {\Omega {r^2}}},
\]
and therefore
\[
Q \to {\rm const.} \ \  {\rm for} \ \ {s}  > -1/2,
\]
where $l_{\Phi,{\rm in}}$ is the angular momentum of $\Phi$ 
at the inner radius $r_{\rm in}$.
Thus, the analysis presented in this paper holds
for ${s}  > -1/2$. For ${s}  < -1/2$, our 
analysis may present a qualitative result.

\section[]{Flicker noise in ADAFs}
The dynamic equations of ADAFs read as follows (\citealp{Narayan94}; \citealp{Kato08}; \citealp{Li09}),
\[
\frac{\partial }{\partial t}(2H\rho) = \frac{1}{2\pi r}\left( {\frac{\partial 
\dot {M}}{\partial r} - {\Phi} } \right),
\]
\[
\rho r\Omega ^2 - \rho r\Omega_{\rm K}^2 - \frac{\partial p}{\partial r} = 0,
\]
\[2H\rho \left[ {\frac{{\partial {\upsilon _\varphi }}}{{\partial t}} + 
\frac{{{\upsilon _{r}}}}{r}\frac{\partial }{{\partial r}}\left( {r{\upsilon _\varphi }} \right)} \right] 
+ \frac{\Phi}{{2\pi {r^2}}}{\Delta}l
 =   \frac{1}{{{r^2}}}\frac{\partial }{{\partial r}}\left( {{2Hr^2}{\tau _{{{r}}\varphi }}} \right),
 \]
\[
\frac{1}{\gamma _2 - 1}\left( {\frac{\partial p}{\partial t}+{\upsilon_r}\frac{\partial p}{\partial r} 
- \gamma _1 \frac{p}{\rho 
}\frac{\partial \rho }{\partial t}- \gamma _1 \frac{p}{\rho}{\upsilon_r}
\frac{\partial \rho }{\partial r}} \right) = - \tau _{r\varphi} \left( 
{r\frac{\partial \Omega }{\partial r}} \right),
\]
where $\gamma_1$ and $\gamma_2$ are the usual 
generalized ratios of the specific heat 
, $\tau_{r\varphi}=-{\alpha}p$, $H/r=\rm const.$, and 
${\Phi}$, ${\Delta}l= {{l_{{\mathop{\Phi}} }} - r\upsilon_\varphi}$ maintain 
the value of the stationary state. The self-similar solution of the above equations is  
\[
\upsilon _{r,0} \propto r^{-1/2},\,\,
\Omega _0 \propto r^{-3/2},\,\,
p_0 \propto r^{ - 5/2 + s},\,\,
\rho _0 \propto r^{ - 3/2 +s}.
\]
We introduce small deviations of the disc parameters from the stationary 
parameters as follows,
\[
\alpha =\alpha_0(1+\overline{\beta}),\;\;\; \overline{\beta}'=r^{s}\overline{\beta},
\]
\[
{\upsilon _{{r}}}=- {a_1}\alpha_0 r{\Omega _{\rm K}}\left( {1 + r^{-s}\upsilon } \right),\;\;\;
\Omega  = {a_2}{\Omega _{\rm K}}\left( {1 + r^{-s}\omega } \right),
\]
\[
p = {a_3}\sqrt r \Omega _{\rm K}^2{r^{s}}\left( {1 + r^{-s}\delta } \right), \;\;\;
\rho  = {a_4}{r^{ - 3/2+s}}\left( {1 + r^{-s}\sigma } \right).
\]
where $a_1$, $a_2$, $a_3$, and $a_4$ are constant.
With the above equations, the evolution 
equations of the perturbed variables are
\[
\frac{1}{\alpha _0 \Omega _{\rm K} }\frac{\partial {\sigma }}{\partial t} =-a_1 
r\frac{\partial \left( {{\upsilon } + {\sigma }} \right)}{\partial r},
\]
\[
a_3 r\frac{d{\delta }}{dr} - \frac{5}{2}a_3 {\delta } - 2a_4 a_2^2 {\omega 
} - a_4 \left( {a_2^2 - 1} \right){\sigma } = 0,
\]
\[
 \frac{1}{\alpha \Omega _{\rm K} }\frac{\partial {\omega }}{\partial t} - a_1 
r\frac{\partial {\omega }}{\partial r} + \frac{a_3 }{a_4 a_2 
}r\frac{\partial {\delta }}{\partial r} - \frac{1}{2}a_1 \left( {1-2s} 
\right){\omega }-\frac{1}{2}a_1 {\upsilon } -
\]
\[  \frac{1}{2}a_1{\sigma } + \frac{a_3 }{2a_4 a_2 }{\delta } 
= \frac{a_3 }{a_4 a_2 }r\frac{\partial 
{\overline{\beta }}'}{\partial r} + \frac{a_3 }{2a_4 a_2 }{\overline{\beta} }',
\]
\[
 \frac{1}{\alpha \Omega _{\rm K} }\frac{\partial {\delta }}{\partial t} - \gamma 
_1 \frac{1}{\alpha \Omega _{\rm K} }\frac{d{\sigma }}{dt} - a_1 r\frac{\partial 
{\delta }}{\partial r} + \gamma _1 a_1 r\frac{\partial {\sigma }}{\partial 
r}- \left( {\gamma _2 - 1} \right)\times \]
\[{a_2}r\frac{{\partial \omega }}{{\partial r}} + \left[ {\frac{5}{2}{a_1} 
+ \left( {s - \frac{3}{2}} \right){\gamma _1}{a_1} + \frac{3}{2}\left( {{\gamma _2} - 
1} \right){a_2}} \right]\delta+
 \]
\[ \left( {\frac{5}{2} - s - \frac{3}{2}\gamma _1 + s\gamma _1 } 
\right)a_1 {\upsilon } -s\gamma _1 a_1 {\sigma }+ \left( {s + \frac{3}{2}} \right) \times
 \]
\[ \left( {\gamma _2 - 1} \right)a_2 {\omega }= - \frac{3}{2}\left( {\gamma _2 - 1} 
\right)a_2 {\beta }'.\]
The accretion rate is
\[
\dot M =  - 4\pi r{\upsilon _{{r}}}\rho H,
\]
and its fluctuating component is 
\[
\dot {m}\left( {r,f} \right) = 4\pi a_1 a_4 \alpha_0 \left( {\frac{H}{r}} 
\right)\left[ {{\upsilon } + {\sigma }} \right].
\]
For $\sqrt {{{\left\langle \bar{\beta}  \right\rangle }^2}}  \propto {r^0}$,
the power spectrum is (see \citealp{Lyubarskii97}, Section 5)
\[
p(f) \propto f^{ - (1 + \frac{4}{3}s)}.
\]

\end{document}